\documentclass[sigconf]{acmart}

\AtBeginDocument{%
  \providecommand\BibTeX{{%
    \normalfont B\kern-0.5em{\scshape i\kern-0.25em b}\kern-0.8em\TeX}}}

\usepackage{enumitem}
\usepackage{courier}
\usepackage{multirow}
\usepackage{amsmath}
\usepackage{breqn}
\usepackage{tikz}
\usepackage{balance}
\usepackage{appendix}
\usepackage{caption}
\usepackage{subcaption}
\usepackage{arydshln}
\usepackage{footnote}
\usepackage{xcolor}
\newcommand{\ie}{\emph{i.e., }}

\newcommand{\etc}{\emph{etc.}}

\copyrightyear{2024} 
\acmYear{2024} 
\setcopyright{acmlicensed}
\acmConference[KDD '24]{Proceedings of the 30th ACM SIGKDD Conference on Knowledge Discovery and Data Mining}{August 25--29, 2024}{Barcelona, Spain}
\acmBooktitle{Proceedings of the 30th ACM SIGKDD Conference on Knowledge Discovery and Data Mining (KDD '24), August 25--29, 2024, Barcelona, Spain}
\acmDOI{10.1145/3637528.3671772}
\acmISBN{979-8-4007-0490-1/24/08}

\newcommand{\method}{\textsc{LARP}}

\DeclareMathOperator*{\argmax}{arg\,max}

\begin{document}

\title{LARP: Language Audio Relational Pre-training for Cold-Start Playlist Continuation} 

\author{Rebecca Salganik}
\affiliation{
    \institution{University of Rochester}
    \city{Rochester}
    \state{NY}
    \country{United States}
}
\email{rsalgani@ur.rochester.edu}
\authornote{Equal contribution.}
\authornote{This work was mainly done while Rebecca was visiting the National University of Singapore.}

\author{Xiaohao Liu}\authornotemark[1]
\affiliation{
    \institution{National University of Singapore}
    \city{Singapore}
    \country{Singapore}
}
\email{xiaohao.liu@u.nus.edu}

\author{Yunshan Ma}
\affiliation{
    \institution{National University of Singapore}
    \city{Singapore}
    \country{Singapore}
}
\email{yunshan.ma@u.nus.edu}
\authornote{Corresponding author.}
\author{Jian Kang}
\affiliation{
    \institution{University of Rochester}
    \city{Rochester}
    \state{NY}
    \country{United States}
}
\email{jian.kang@rochester.edu}

\author{Tat-Seng Chua}
\affiliation{
    \institution{National University of Singapore}
    \city{Singapore}
    \country{Singapore}
}
\email{dcscts@nus.edu.sg}

\renewcommand{\shortauthors}{Rebecca Salganik, Xiaohao Liu, Yunshan Ma, Jian Kang, \& Tat-Seng Chua}

\begin{abstract}

As online music consumption increasingly shifts towards playlist-based listening, the task of \textit{playlist continuation}, in which an algorithm suggests songs to extend a playlist in a personalized and musically cohesive manner, has become vital to the success of music streaming services. Currently, many existing playlist continuation approaches rely on collaborative filtering methods to perform their recommendations. However, such methods will struggle to recommend songs that lack interaction data, an issue known as the \textit{cold-start} problem. Current approaches to this challenge design complex mechanisms for extracting relational signals from sparse collaborative signals and integrating them into content representations. However, these approaches leave content representation learning out of scope and utilize frozen, pre-trained content models that may not be aligned with the distribution or format of a specific musical setting. Furthermore, even the musical state-of-the-art content modules are either (1) incompatible with the cold-start setting or (2) unable to effectively integrate cross-modal and relational signals. In this paper, we introduce \method, a multi-modal cold-start playlist continuation model, to effectively overcome these limitations. \method~is a three-stage contrastive learning framework that integrates both multi-modal and relational signals into its learned representations. Our framework uses increasing stages of task-specific abstraction: within-track (language-audio) contrastive loss, track-track contrastive loss, and track-playlist contrastive loss. Experimental results on two publicly available datasets demonstrate the efficacy of \method~over uni-modal and multi-modal models for playlist continuation in a cold-start setting. Finally, this work pioneers the perspective of addressing cold-start recommendation via relational representation learning. Code and dataset 
are released at: \href{https://github.com/Rsalganik1123/LARP}{https://github.com/Rsalganik1123/LARP}.
\end{abstract}




\maketitle

\section{Introduction} 

In recent years, the rise of music streaming platforms has created significant changes in the orientation of music listening practices \cite{schedl_current_2018}. Most notably, the transition from physical to digitized music has manifested itself in the dominance of playlist-based listening over, the previously prevalent, album-based listening \cite{bonnin_automated_2014, drott_why_2018, prey_locating_2020, plist_popularity}. In the general sense, a playlist can be considered as a collection of tracks (songs\footnote{Track and song are used interchangeably in this paper.}), intended to be listened to in sequential order and unified under some underlying theme such as genre, mood, activity, aesthetic, or other cohesive factor \cite{zamani_analysis_2019, bontempelli_flow_2022, dias_user_2014}. This radical shift in consumption habits has highlighted the significance of \textit{playlist continuation}, a task which requires an algorithmic curatorial system to augment a list of seed tracks with musically cohesive suggestions to complete a playlist \cite{bonnin_automated_2014, zamani_analysis_2019}. 

 \begin{figure} 
    \centering
    \includegraphics[width=0.95\linewidth]{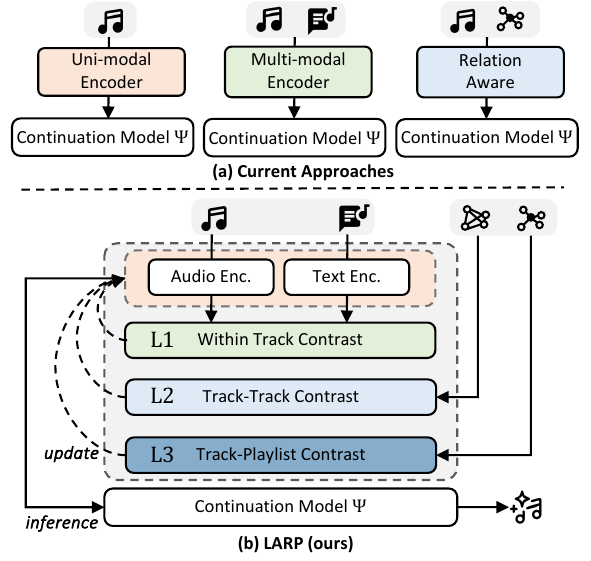}
    \vspace{-3mm}
    \caption{Comparison of cold-start approaches. Current cold-start learning methods (top) delegate content representation to a pre-trained model and focus on relational extraction. However, LARP (bottom) simultaneously focuses on relational and cross-modal learning, addressing both the limitations of previous content models and enhancing the performance of downstream cold-start frameworks.}
    \label{fig:method_comparisons}
    \vspace{-1em}
\end{figure}

To date, a large portion of approaches for playlist continuation rely on collaborative filtering (CF), which leverages user-item interactions to learn user and item representations. However, pure collaborative filtering hardly addresses the \textit{cold-start problem}, where a lack of interactive data hinders the ability of a system to generate high-quality item or user representations and offer meaningful suggestions~\cite{wenjie_EQUAL_23, barkan_cb2cf_2019, wei_clcrec_2021, li_from_2019, saveski_item_2014, vartak_meta_2017, volkovs_dropoutnet_2017, zhu_recommendation_2020, zhang_addressing_2014}. Broadly, solutions to the cold-start problem require methods which infuse a set of content representations with informative relational signals, enabling a model to later generalize to unseen users and/or items. Existing works approach the cold-start problem by designing complex systems that constrain \cite{li_from_2019, saveski_item_2014}, align \cite{wu_m2eu_2023}, or reconstruct \cite{du_how_2020, volkovs_dropoutnet_2017, shi_adaptive_2019, barkan_cb2cf_2019} relational signals and use them to enhance a set of pre-computed content representations.  Thus, central to the success of these methods is the quality of the representations generated by their selected content module. Crucially, these approaches leave the generation of their content representations out of scope, delegating feature extraction to a frozen, pre-trained content extraction module. Meanwhile,  music and audio representation learning has presented a wealth of content-based feature extractors which can serve as content-backbones in the aforementioned cold-start approaches. These methods encompass uni-modal \cite{mccallum_supervised_2022, park_representation_2018, kazakos_slow-fast_2021, chen_htsat_2022, kong_panns_2020, spijkervet_contrastive_2021, won_semisupervised_2021, tagliascchi_pretraining_2020}, language-audio \cite{wu_largescale_2023, manco_learning_2022, huang_mulan_2022, gardner_llark_2023}, and relational approaches \cite{jimenez_pretraining_2023, huang_largescale_2020, ferraro_contrastive_2023}.

The state-of-the-art (SOTA) in music representation learning have several key limitations with regard to the task of playlist continuation in a cold-start setting: either they (1) rely on collaborative filtering approaches to extract relational signals and are therefore incompatible with the cold-start setting due to a reliance on interactive data being present during inference \textbf{or} (2) they are unable to effectively integrate cross-modal and relational signals. Given our setting, in which inference is performed over a set of unseen playlists and tracks, there is an urgent need for comprehensive feature extraction methods that combine both relational and multi-modal learning but do not rely on collaborative filtering methods to do so.


In this work, we present a fundamental paradigm shift, transferring the focus of cold-start learning away from the complicated extraction of realtional signals and towards the training of relational awareness \textit{within} a multi-modal content module. We argue that the quality of the representations learned by a content module poses a significant bottleneck for cold-start learning. To this end, we introduce \method, a \underline{L}anguage \underline{A}udio \underline{R}elational \underline{P}re-training model, which effectively overcomes the limitations of previous SOTA music representation learning methods. LARP approaches the task of playlist continuation through a three-layered contrastive learning framework that integrates both multi-modal and relational signals into its learned representations with increasing layers of task-specific abstraction. 
We extend the canonical contrastive learning framework and introduce the concept of multi-stage contrastive learning by designing cross-modal contrastive loss functions for within-track, track-track and track-playlist pairings. Our approach centers around three key loss functions: (L1) \textit{Within-Track Contrastive Loss}, (L2) \textit{Track-Track Contrastive Loss}, and (L3) \textit{Track-Playlist Contrastive Loss}, visualized in Figure \ref{fig:method_comparisons}. First (L1), individual tracks are passed through an language-audio encoder, which is trained by enforcing alignment between their respective textual and audio embeddings. Then (L2) these representations are further refined by aligning the representations between pairs of tracks which share a parent playlist. In the final layer (L3), we generate representations for playlists enforcing alignment between a subset of tracks within a playlist. In order to show the superiority of our methodology, we augment two publicly available music datasets to include audio samples and meta-data based textual annotations as multi-modal sources of information. 
Extensive experiments show that our method outperforms both uni-modal and multi-modal models to achieve impressive results on the cold-start playlist continuation task.

In summary, the main contributions of this paper are as follows: 
\begin{itemize}[leftmargin=*, noitemsep,topsep=0pt]
    \item \textbf{Problem Definition.} We argue that the quality of the representations learned by a content module poses a significant bottleneck for cold-start learning and address this challenge through relational pre-training.
    \item \textbf{Model Design.} We propose \method, which is the first multi-modal relational pre-training framework that uses multi-stage contrastive learning to generate embeddings.
    
    \item \textbf{Experimental Results.} We augment two canonical music recommendation datasets with audio waveforms and their meta-data based textual annotation counterparts. Then, through extensive experimentation and analysis, we demonstrate the effectiveness of our method in addressing the needs of cold-start playlist continuation 
\end{itemize}

\section{Problem Formulation}\label{sec:prob_def}

Given a set of tracks, $\mathcal{S} = \{s_1, s_2, \cdots, s_N\}$, each track $s$ includes an audio input $a_s$ and a text input $t_s$, which could be the description about its genre, artist, or title \etc~ (For the exact formulations of our own audio and textual inputs, please see Section \ref{sec:experiments}). We define a playlist as a \textit{set}~\footnote{Please note: in our setting we do not consider the sequential nature of playlists. Rather, we leave this characteristic for future work and simplify a playlist to be an unordered set of tracks.} of tracks, denoted as $p = \{s_1, s_2, \cdots, s_{\lvert p \rvert}\}$, where $\lvert p \rvert$ is the size of the playlist. Typically, we are given a collection of playlists as the training set, denoted as $\mathcal{P}=\{p_1, p_2, \cdots, p_M\}$ and a collection of playlists for testing, denoted as $\mathcal{\bar{P}}=\{\bar{p}_{M+1}, \bar{p}_{M+2}, \cdots, \bar{p}_{M+\bar{M}}\}$, where $M$ is the size of the training set and $\bar{M}$ is the size of the testing set. Associated with each playlist set is a collection of tracks that correspond to the playlists in the training set, $\mathcal{S}=\{s_1, s_2, \cdots, s_N\}$, and testing set, $\mathcal{\bar{S}}=\{\bar{s}_{N+1}, \bar{s}_{N+2}, \cdots, \bar{s}_{N + \bar{N}}\}$. The interactions between these tracks and their parent playlists, are represented by a bipartite graph $\mathbf{R}_{M \times N}=\{r_{ps}|p\in{\mathcal{P}},s\in{\mathcal{S}}\}$, where $r_{ps}=1$ if $s \in p$ and $r_{ps}=0$ otherwise. Based on the track-playlist affiliation relations in $\mathbf{R}$, we further define a track-track co-occurrence relation between every pair of tracks which share a parent playlist, forming a homogeneous graph $\mathbf{O}_{N \times N}=\{o_{s_{i}s_{j}}|s_{i}\in{\mathcal{S}},s_{j}\in{\mathcal{S}}\}$, $o_{s_{i}s_{j}}=1$ if the tracks $s_{i}$ and $s_{j}$ belong to the same playlist, and $o_{s_{i}s_{j}}=0$ otherwise. 

\begin{itemize}[leftmargin=*, noitemsep,topsep=0pt]
    \item \textbf{Playlist Continuation:} Given a test playlist, $\bar{p}_q \subset \bar{p} \in \mathcal{\bar{P}}$, consisting of $q$ seed test tracks, $J = \{ \bar{s}_j: \bar{s}_j \in \bar{p}_q\}$ such that $\lvert J \rvert < \lvert \bar{p} \rvert$, the task of playlist continuation aims to predict the missing tracks $\bar{s} \in \bar{p} \setminus J$. 
    
    \item \textbf{Cold-start Playlist Continuation:} In this work, we define the cold-start setting to mean that there is no overlap between the tracks and playlists in the training and testing sets. Thus, $\mathcal{S} \cap \mathcal{\bar{S}}= \emptyset $ and $\mathcal{P} \cap \mathcal{\bar{P}} = \emptyset$ \footnote{Please note: that throughout this paper, we use $\bar{s} \in \mathcal{\bar{S}}, ~~\bar{p} \in \mathcal{\bar{P}}$ to distinguish playlists and tracks in the test set.}. Here, the set of training tracks, $\mathcal{S}$ has size, $N$ and the set of testing tracks, $\mathcal{\bar{S}}=\{\bar{s}_{N+1}, \bar{s}_{N+2},\cdots, \bar{s}_{N+\bar{N}}\}$, of size, $\bar{N}$, is associated with the testing playlists, $\mathcal{\bar{P}}$. 
    
\end{itemize}

\section{Methodology}

We introduce LARP, a novel language audio relational pre-training method, targeted at extracting track representations that are effective and generalizable to various methods for cold-start playlist continuation. We first present the model architecture of LARP, followed by the multi-stage pre-training process, and finally we describe how to employ LARP for cold-start playlist continuation in tandem with several canonical cold-start frameworks.

\begin{figure*}[h]
     \centering
     \includegraphics[width=0.99\linewidth]{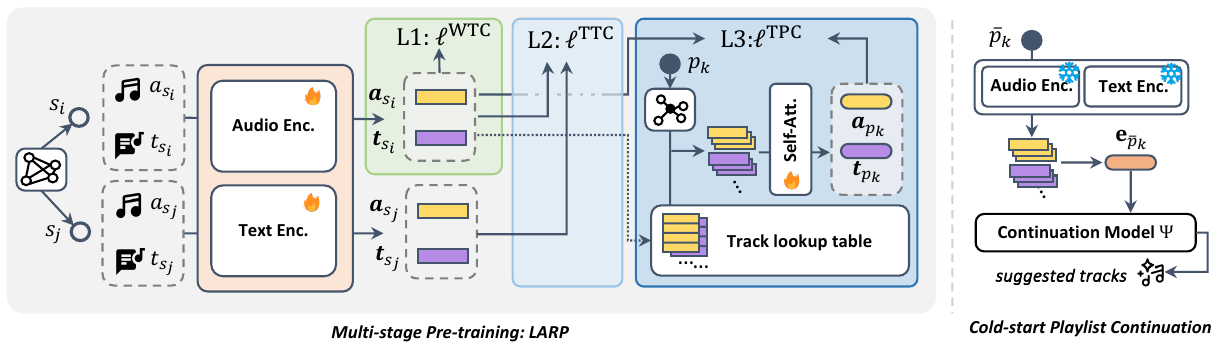}
     \vspace{-3mm}
     \caption{Overview of LARP. During training (left), LARP uses three stages of contrastive loss. In the first stage of training, LARP performs Within-Track Contrastive Loss (WTC) on the text, audio input from a single track. In the second stage of training, \method~performs Track-Track Contrastive Loss (TTC) on the text, audio pairs of co-occurring songs. In the final stage, \method~performs Track-Playlist Contrastive Loss (TPC) on the text, audio pair of a playlist and its child track. During testing (right), LARP is used to generate track embeddings which are pooled to create playlist representations and used as content-input to a downstream cold-start framework to perform playlist continuation.}
        \label{fig:full_LARP}
\end{figure*}


\subsection{Model Architecture}\label{sec:model_architecture}

In order to generate audio and text embeddings for each track in our track dataset, we design a multi-modal architecture that is based on the BLIP framework \cite{li_blip_20222}. As shown in Figure \ref{fig:full_LARP}, our method consists of two uni-modal encoders: one for audio and the other for text. For more extensive details on the formatting of out audio and textual inputs, please see Section \ref{sec:experiments}.

Following a similar architecture with \cite{wu_largescale_2023}, we select \emph{HT-SAT} \cite{chen_htsat_2022}, a spectral transformer, for our audio encoding and Bert \cite{devlin_pretraining_2019}, a language transformer, for our textual encoding. In our experimentation, we tried several other audio backbones such as AST \cite{gong_ast_2021} and PANN \cite{kong_panns_2020} however each of these had specific drawbacks. First, similarly to \cite{wu_largescale_2023}, we found that the convolutional architecture of the PANN \cite{kong_panns_2020} model did not achieve high performance on the audio encoding task. Second, due to the computational complexity of the AST \cite{gong_ast_2021} model, we found that the number of parameters significantly limited the batch size that we were able to work with, thus deteriorating performance. As such, we use the HT-SAT, a lighter modified version of AST \cite{gong_ast_2021}, to generated our audio representations and leave further backbone analysis for future works. 
Thus, given a track,  $s_i$, each uni-modal encoder outputs a representation associated with the audio and textual caption of the track, $\mathbf{a}_{s_i} \in \mathbb{R}^{1 \times 768}, \mathbf{t}_{s_i} \in \mathbb{R}^{1 \times 768}$, respectively. These are then projected into a unified space where they are represented as $\mathbf{a}_{s_i} \in \mathbb{R}^{1 \times 256}, \mathbf{t}_{s_i} \in \mathbb{R}^{1 \times 256}$. 
In addition to the uni-modal encoders and their cross-modal alignment layers, our architecture includes a momentum encoder which is used to store track representations from previous epochs that can be used during the contrastive loss. This design choice is motivated by the substantial computational cost of generating and optimizing  track representations, and we follow the previous practice in BLIP~\cite{li_blip_20222} for our implementation of the momentum encoder. This encoder can be seen as a queue which caches representations of tracks from the forward pass of each batch. Each time, a new representation is generated, it is integrated into the queue using a first-in first-out manner. As we explain in the subsequent Section ~\ref{sec:training_objective}, this cache is used for sampling negative examples during the optimization of the WTC and TTC losses.

\subsection{Multi-stage Pre-training}\label{sec:training_objective}
In order to train our model, we design three contrastive loss variations WTC, TTC, and TPC, each associated with an increasing level of task abstraction. The three losses are integrated via three consecutive stages of training, in which the weights from a previous stage are transferred to the next. We begin by detailing them individually and then explain the paradigm used for their combination. For a visualization, please see Figure~\ref{fig:full_LARP}. 
\subsubsection{\textbf{(L1) WTC: Within-Track Contrastive Loss}}
We borrow the ITC, \textit{Image-Text Contrast},  loss formulation from the original BLIP \cite{li_blip_20222} framework and use it for \textit{Within-Track Contrastive} loss. The motivation behind this loss is grounded in the improved generalization that contrastive loss can provide \cite{wu_largescale_2023, mccallum_supervised_2022, jimenez_pretraining_2023}. Thus, for our first layer of representation learning, we formally define this loss as: 
\begin{equation}\label{wtc}
   \ell_{s_i}^{\text{WTC}} = \text{Contrast}(\mathbf{a}_{s_i}, \mathbf{t}_{s_i}),
\end{equation} 
where $s_i$ is a track in the training set and $\mathbf{a}_{s_i}, \mathbf{t}_{s_i}$ are the audio and textual embeddings which are extracted by our multi-modal encoder $\Phi(\cdot)$. Here, the $\text{Contrast}(\cdot, \cdot)$ function follows the formulation proposed by~\cite{junnan_align_2021}. More concretely, for an input track $s_i$, we calculate the softmax-normalized audio-to-text (a2t) and text-to-audio (t2a) similarity as:
\begin{equation}~\label{eq:InfoNCE}
\begin{split}
     \hat{y}_i^{\text{a2t}} = \frac{\text{exp}(\operatorname{sim}(\mathbf{a}_{s_i}, \mathbf{t}_{s_i}) \mathbin{/} \tau)} {\sum_{j=1}^{N} \text{exp}(\operatorname{sim}(\textbf{a}_{s_i}, \textbf{t}_{s_j}) \mathbin{/} \tau)}, \\
     \hat{y}_i^{\text{t2a}} = \frac{\text{exp}(\operatorname{sim}(\mathbf{t}_{s_i}, \mathbf{a}_{s_i}) \mathbin{/} \tau)}  {\sum_{j=1}^{N} \text{exp}(\operatorname{sim}(\textbf{t}_{s_i}, \textbf{a}_{s_j}) \mathbin{/} \tau)}, 
\end{split}
\end{equation}
where sim($\cdot$) is the cosine similarity function, $\tau$ is a temperature hyper-parameter, and $s_j$ are the negative samples, sampled from the momentum queue. Then, given the one-hot ground-truth similarity vectors, $y_i^{\text{a2t}}, y_i^{\text{t2a}} \in \{0,1\}$, the contrastive loss is defined as the cross entropy $\text{CE}(\cdot)$ between the predicted similarities $\hat{y}_i^{\text{a2t}}, \hat{y}_i^{\text{t2a}}$ and the ground truths $y_i^{\text{a2t}}, y_i^{\text{t2a}}$. Such that: 
\begin{equation} \label{wtc_def}
     \text{Contrast}(\mathbf{a}_{s_i}, \mathbf{t}_{s_i}) =  \frac{1}{2} [\text{CE}(y_i^{\text{a2t}}, \hat{y}_i^{\text{a2t}}) + \text{CE}(y_i^{\text{t2a}}, \hat{y}_i^{\text{t2a}} )].
\end{equation}

\subsubsection{\textbf{(L2) TTC: Track-Track Contrastive Loss.}}
In the second stage of training, we refine the representations learned from the WTC loss by applying contrastive loss to align the modalities between pairs of tracks. We refer to this loss as TTC, or \textit{Track-Track Contrastive} loss. The purpose of this loss is to integrate relational signals into the learned representations of individual songs. This layer of contrast allows the model to learn essential information that can be used to unify songs in both the language and audio representational spaces. For two tracks $s_i$ and $s_j$, which share a parent playlist $p_k$, we can define this loss as the following: 
\begin{equation}
     \ell_{s_i,s_j}^{\text{TTC}} = \frac{1}{2} [\text{Contrast}(\textbf{a}_{s_i}, \textbf{t}_{s_j}) + \text{Contrast}(\textbf{t}_{s_i}, \textbf{a}_{s_j})],  
\end{equation}
where the $\text{Contrast}(\cdot,\cdot)$ loss is similar with Equation \ref{wtc}, but, we change the contrast pair from single track to two different tracks. The negative samples are also looked up from the momentum encoder. For the full equation, please refer to Appendix~\ref{ap:loss}. 

\subsubsection{\textbf{(L3) TPC: Track-Playlist Contrastive Loss.}}
Finally, given that the ultimate goal of automatic playlist continuation requires matching between playlists and tracks, we build on the previous loss function to develop a contrastive loss between a track, $s_i$, and its parent playlist, $p_k$. The purpose of this loss is to create alignment between clusters of tracks that share a parent playlist, thus improving the neighbourhood awareness of a learned representation. In order to achieve this level of alignment, we design the TPC, short for \textit{Track-Playlist Contrastive}, loss. Thus, given a central track, $s_i$, its parent playlist, $p_k$, and a set of neighbour tracks, $J = \{ s_j : s_j \in p_k \setminus s_i\}$, we define the loss as follows: 
\begin{equation}\label{eq:final_loss}
     \ell_{p_k,s_i}^{\text{TPC}} = \frac{1}{2} [\text{Contrast}(\textbf{a}_{p_k}, \textbf{t}_{s_i}) + \text{Contrast}(\textbf{t}_{p_k}, \textbf{a}_{s_i})],  
\end{equation} where $\mathbf{a}_{p_k}, \mathbf{t}_{p_k} \in \mathbb{R}^{1 \times 256}$ are the playlist audio and text representations, respectively which are obtained by aggregating over a series of $J$ seed tracks associated with each playlist using the Equation \ref{eq:fusion} below. 

\noindent \textbf{TPC-\textit{fusion}.} We experiment with several aggregation functions in order to generate a playlist embeddings, $(\mathbf{a}_{p_k}, \mathbf{t}_{p_k})$. While a simple average would be the most computationally efficient, in practice, this method has some limitations. Primarily, it is unable to account for outliers in a track collection. For example, in a diverse playlist, it is difficult to guarantee that 
all of the selected tracks will be aligned within the same theme. Thus, using a simple average as an aggregator can cause the noise in the learned representation of a playlist, and significantly mislead the optimization of the TPC objective. To address this problem, we propose a \textit{fusion} module to guide  the playlist representation during training~\footnote{Please note that the \textit{fusion} module is not used to generate playlist representations during testing, \ie playlist continuation, only training.}. Specifically, we employ one simple self-attention layer on top of the set of track representations to automatically weigh the tracks within a playlist. Thereby, the outlier or noisy tracks would be assigned lower weights when forming the playlist representation. Formally, given a playlist $p_k$, its representation is calculated by:   
\begin{equation} \label{eq:fusion}
\begin{aligned}
    \textbf{a}_{p_k} &= \text{Self-Att}(\{\textbf{a}'_{s_j}: s_j \in J\}), \\ \textbf{t}_{p_k} &= \text{Self-Att}(\{\textbf{t}'_{s_j}: s_j \in J\}),
\end{aligned}
\end{equation} 
where $J = \{ s_j : s_j \in p_k \setminus s_i\}$, \text{Self-Att($\{\cdots\}$)} denotes a one-layer self-attention network, $\textbf{a}_{p_k}$ and $\textbf{t}_{p_k}$ are the audio and text representations for $p_k$, and $\textbf{a}'_{s_j}$ and $\textbf{t}'_{s_j}$ are the audio and text representations of tracks within $p_k$ (excluding $s_i$), which are drawn from the non-differentiable track representation lookup dictionary $\mathbf{A}_{\Tilde{S}}$ and $\mathbf{T}_{\Tilde{S}}$. We note that this track representation lookup table differs from the momentum encoder used in the previous WTC and TTC losses. This is because the momentum encoder has limited capacity and cannot store all of the tracks, which could skew the results of the playlist aggregation. To address this issue, we leverage two non-differentiable lookup tables to cache the latest audio and text representations for all the tracks during training, denoted as $\mathbf{A}_{\Tilde{S}} \in \mathbb{R}^{N \times 256}$ and $\mathbf{T}_{\Tilde{S}} \in \mathbb{R}^{N \times 256}$. Both tables are updated during training such that when a track appears in a batch their representations in $\mathbf{A}_{\Tilde{S}}$ and $\mathbf{T}_{\Tilde{S}}$ will be replaced by the new representations generated by the forward pass. For a more detailed breakdown to this loss function and its calculation procedures, please refer to Appendix~\ref{ap:loss}.



\noindent \textbf{Multi-stage Training.}
We train the model in three stages, using early stopping to achieve the optimal performance for each loss and before proceeding to the following stage. The weights from each intermediate stage are transferred between stages and the model remains unchanged. Only the loss function is modified to integrate each increasingly abstract loss. Thus, for a given track $s_i$ and its affiliated playlist $p_k$, and \textit{neighbour set}, $\{s_j: p_k = p(s_j), s_j \neq s_i\}$, we define the stages as: 
\begin{align}
\text{Stage 1:}~\ell_1 &= ~\ell_{s_i}^{\text{WTC}}; & \\
\text{Stage 2:}~\ell_2 &= ~ \ell_{s_i}^{\text{WTC}} +  \ell_{s_i, s_j}^{\text{TTC}}; & \\ 
\text{Stage 3:}~\ell_3 &= ~ \ell_{s_i}^{\text{WTC}} + \ell_{s_i, s_j}^{\text{TTC}} + \ell_{p_k, s_i}^{\text{TPC}}. &
\end{align}
Currently, we just use a simple average over the three losses, which already demonstrates excellent performance. We leave more complex balances between the three loss functions for future work.

\subsection{Cold-start Playlist Continuation}\label{sec:rec}

After the pre-training stage, we obtain the model $\Phi(\cdot)$ for cold-start playlist continuation task. 
Specifically, we first generate the text and audio representations: 
\begin{equation} \label{eq:test_rep_extraction}
    \textbf{a}_{s_i}, \textbf{t}_{s_i} = \Phi([a_{s_i}, t_{s_i}]), 
\end{equation}
where $a_{s_i}$ and $t_{s_i}$ stand for the audio and text input of the track $s_i$, respectively. 
Analogous to the pre-training, we employ a simple average pooling to obtain its unified representation $\mathbf{e}_{s_i} \in \mathbb{R}^{1 \times 256}$, formally represented as:
\begin{equation} \label{eq:modal_avg}
    \mathbf{e}_{s_i} = \text{mean} (\textbf{a}_{s_i}, \textbf{t}_{s_i}).
\end{equation}

Thus, we can obtain the embedding table for all tracks in training set $\mathcal{S}$ as $\mathbf{E}_{\mathcal{S}} \in \mathbb{R}^{N \times 256}$, and testing set $\mathcal{\bar{S}}$ as $\mathbf{E}_{\mathcal{\bar{S}}} \in \mathbb{R}^{\bar{N} \times 256}$.

Following the typical setting in playlist continuation \cite{schedl_current_2018,zamani_analysis_2019}, we generate the playlist representations $\mathbf{e}_{p}$ and $\mathbf{e}_{\bar{p}_q}$ by simply average pooling its included tracks' representations, denoted as:
\begin{equation} \label{eq:plist_gen}
   \mathbf{e}_{p} = \text{mean}(\{\mathbf{e}_{s_i}: s_i \in p\}),\quad 
   \mathbf{e}_{\bar{p}_q} = \text{mean}(\{\mathbf{e}_{\bar{s}_i}: \bar{s}_i \in \bar{p}_q\}),
\end{equation}
where $\mathbf{e}_{p}$ and $\mathbf{e}_{\bar{p}_q}$ are used for training and testing, respectively. 
And the cold-start playlist continuation task aims to predict the missing tracks $\bar{s}\in \bar{p} \setminus \bar{p}_q$.

In order to showcase the ability of LARP to learn effective and generalizable representations, we integrate the representations generated by Equations \ref{eq:test_rep_extraction} and \ref{eq:plist_gen} 
with three cold-start playlist continuation methods: ItemKNN~\cite{sarwar_item_2001}, DropoutNet~\cite{volkovs_dropoutnet_2017}, and CLCRec~\cite{wei_clcrec_2021}. It should be noted that these methods are originally designed for cold-start recommendation and we adapt them to the task of cold-start playlist continuation (please see \ref{ap:hp_cold_start} for more details).
\begin{itemize}[leftmargin=0.5cm, itemindent=0cm, noitemsep, topsep=0pt]
\item \textbf{ItemKNN}~\cite{sarwar_item_2001} is a parameter-free approach which relies on the cosine similarity between the input partial playlist $\mathbf{e}_{\bar{p}_q}$ and all the candidate tracks $\mathbf{E}_{\mathcal{\bar{S}}}$, to find the $K$ most similar tracks. 
\item \textbf{DropoutNet}~\cite{volkovs_dropoutnet_2017} is designed to treat the cold-start as a robustness task in which sparse interactive signals are reconstructed using content features. 
It adopts WMF~\cite{koren_collaborative_2008} to generate preference representations for playlists and tracks, and randomly applies dropout to a training set of preference-content concatenations, which are fed into a DNN model for score predictions. 

\item \textbf{CLCRec}~\cite{wei_clcrec_2021} is a contrastive learning framework that reconstructs collaborative signals from the provided content embeddings. And it also uses inner product to compute the similarities, then to find the $K$ possible tracks in the  $\bar{p} \setminus \bar{p}_q$.
\end{itemize}



\section{Experiments}
To guide our analysis, we center our experimentation around answering the following research questions: 
\begin{itemize}[leftmargin=0.9cm, itemindent=0cm, noitemsep,topsep=0pt]
    \item [\textbf{RQ1.}] How does our proposed method perform in comparison with our selected benchmarks on the task of playlist continuation? 
    \item [\textbf{RQ2.}] How does each block of our proposed framework contribute to the final performance of our method? 
  
    \item [\textbf{RQ3.}] What are the key properties of our method that make it well suited to the task of cold-start playlist continuation? 
    
\end{itemize}

\vspace{-.1in}
\subsection{Experimental Settings} \label{sec:experiments}

\subsubsection{Datasets}

We evaluate LARP on The Million Playlist Dataset (MPD)~\cite{MPD} and the Last-FM (LFM)~\cite{LFM} datasets (Table~\ref{table:dataset_stats}). Our motivation for selecting these datasets is grounded in two key requirements of our setting: (1) direct access to audio samples and metadata, and (2) inclusion of track-based interactions. While there are several playlist datasets available \cite{mcfee_hypergraph_2012,zangerle_nowplaying_2014, turrin_30music_2015, ferraro_melon_2021}, none of them would give us access to the audio samples. And, even though the Melon dataset \cite{ferraro_melon_2021} has pre-computed spectrograms, the sampling frequency and time duration of these spectrograms are not compatible with the needs of our backbone audio extraction module.

We define our own procedure for augmenting both MPD and LFM to include the language-audio pairs necessary for our multi-modal pre-training methodology. For generating the audio samples, we scrape all the associated MP3s and truncate them into 10 second clips sampled randomly from the track (excluding the first and last 20 seconds). Then, to meet the size requirements of our backbone audio model, HTS-AT \cite{chen_htsat_2022}, we truncate each MP3 file into 10 second sound clips and upsample the audio quality to be 44.8kHz. To generate the captions, we use the associated Spotify IDs to collect the track name, artist name, and album name. The captions are then formulated using the following structure: The track \texttt{<track name>} by \texttt{<artist name>} on album \texttt{<album name>}. We leave the exploration of further captioning strategies for future work.

Finally, we acknowledge that LFM is not explicitly designed for playlist continuation, so we make several modifications to align it with our task. We draw inspiration from the notion of Weekly Discovery. This service, which is available on many popular streaming platforms \cite{stanisljevic_impact_2020, bonnin_automated_2014}, involves the creation of a playlist on the basis of the aggregated listening habits over a user on a weekly basis. Unlike static playlist continuation, in which seed tracks are collected from the user inputs, Weekly Discovery uses implicit feedback from users to identify the seed tracks used for recommendation. Thus, to create a playlist continuation dataset from the user-track interaction data provided by LFM, we perform the following steps: 
\begin{enumerate}[leftmargin=0.5cm, itemindent=0cm, noitemsep, topsep=0pt]
    \item We randomly select a series of users for each split of training, validation, and testing. 
    \item We treat each user as a `playlist' and aggregate all the songs that they have listened to into a single collection. 
    \item We iteratively remove songs which overlap between the split pools (i.e., songs in the training set are removed from the subsequent validation and test sets). 
    \item To construct our test set, we select a contiguous subset of interactions for each user/playlist to contain more than 30 songs and less than 100 for standardization between MPD and LFM. 
\end{enumerate}

\subsubsection{Baseline methods.}
We compare LARP with four state-of-the-art music representation learning baselines, including Jukebox~\cite{dhariwal_jukebox_2020}, MULE~\cite{mccallum_supervised_2022}, CO-Playlist~\cite{jimenez_pretraining_2023}, and CLAP~\cite{wu_largescale_2023}. Please note that, due to the cold-start nature of our dataset design, we do not include any collaborative filtering baselines, which require interactive data for representation learning. 

\begin{itemize}[leftmargin=0.5cm, itemindent=0cm, noitemsep, topsep=0pt]
    
    \item \textbf{Jukebox}~\cite{dhariwal_jukebox_2020} is a generative music model with state-of-the-art performance in various music understanding tasks~\cite{castellon_codified_2021}.
    
    \item \textbf{MULE}~\cite{mccallum_supervised_2022} is an audio representation learning model based on convolutional neural network (CNN) \cite{kong_panns_2020}.  
   
    \item \textbf{CO-Playlist}~\cite{jimenez_pretraining_2023} is a contrastive learning model, in which positive pairs are selected through playlist-track interaction data. 
    
    \item \textbf{CLAP-PT/ CLAP-FT}~\cite{wu_largescale_2023}~\footnote{There is a concurrent work ~\cite{CLAP-ICASSP23-microsoft} that has the same name and similar idea with CLAP~\cite{wu_largescale_2023} that we implemented in this paper.} is a multi-modal model. For the pre-trained (PT) setting, we use the pre-trained model to generate representations. For the fine-tuned (FT) setting, we fine-tune before generating representations.  
\end{itemize}

\subsubsection{Hyper-parameter settings and reproducibility.}
We train LARP in a distributed setting with two Nvidia A40 GPUs, each with 48GB of memory. We train for $45$ epochs with early stopping that terminates after $5$ epochs without improvement. We use a batch size of 50 and use the Adam optimizer with $\beta_1 = 0.9$ and $\beta_2=0.99$. Our learning rate of $1e^{-4}$ is controlled by a cosine scheduler after a $3000$ step warm up period. Our final results are achieved with an embedding size of $256$. For our temperature parameter we use $\tau = 0.07$. Our momentum encoder has a queue of size $57600$ and a momentum parameter of $m=0.995$.  For the hyperparameters used in our cold-start frameworks, please see Appendix \ref{ap:hp_cold_start}.

\begin{table}[t]
\centering

\caption{Statistics of datasets with number of tracks (\#S), number of playlists (\#P), and Homogeneity (Homo.). Homogeneity is defined as the average pairwise cosine similarity among the audio embeddings of the tracks in a playlist.}

\footnotesize
\begin{tabular}{l cccccc }
\toprule
\multirow{2.4}{*}{Dataset} &
\multicolumn{1}{c}{Training} & 
\multicolumn{2}{c}{Evaluation} & 
\multicolumn{3}{c}{Relevant Stats}  \\ 
\cmidrule(lr){2-2} \cmidrule(lr){3-4}\cmidrule(lr){5-7}
& \#S & \#P & \#S & Avg \#S/P & Avg \#P/S & Homo.\\
\midrule
MPD & 98,552 &100 & 6,978  & 98.12 &1.41 &0.60\\
LFM &87,558 & 1,000 & 17,439  & 91.04 &4.63 &0.56\\
\bottomrule
\end{tabular}
\label{table:dataset_stats}
\end{table}

\subsection{Overall Performance Comparison (RQ1)}

\setlength\dashlinedash{0.2pt}
\setlength\dashlinegap{1.5pt}
\setlength\arrayrulewidth{0.3pt}

\begin{table*} [h]
\centering
\caption{Effectiveness results on Recall@K (R@K) and NDCG@K (N@K). Best results in \textbf{bold} and second best with \underline{underscore}.}

\label{table:overall_performance_new}
\resizebox{\linewidth}{!}{
\begin{tabular}{ll cccccc cccccc }
\toprule 

\multirow{2.8}{*}{\textbf{Rec. Model}} &
\multirow{2.8}{*}{\textbf{Encoder}} & 
\multicolumn{6}{c}{\textbf{MPD Dataset}} & 
\multicolumn{6}{c}{\textbf{LFM Dataset}} \\
\cmidrule(lr){3-8} \cmidrule(lr){9-14} &&
R@10 & N@10 & R@20 & N@20 & R@40 & N@40 & R@10 & N@10 & R@20 & N@20 & R@40 & N@40  \\
\midrule

\multirow{7.4}{*}{\textbf{ItemKNN} \cite{sarwar_item_2001}} 

& Jukebox \cite{dhariwal_jukebox_2020}&
0.0012 & 0.0045 & 
0.0025 & 0.0051 &
0.0055 & 0.0059 & 
0.0005 & 0.0017  & 
0.0008 & 0.0015  & 
0.0021 & 0.0019  \\ 

& MULE \cite{mccallum_supervised_2022}&
0.0156&	0.0798 &
0.0286 & 0.0738 &
0.0457 & 0.0609 &
\underline{0.0036} & \underline{0.0140}  & 
0.0066 & 0.0131  & 
0.0121 & 0.0133 \\

& CO-Playlist \cite{jimenez_pretraining_2023}& 
0.0012 & 0.0058 & 
0.0022& 0.0056 &
0.0067 & 0.0076 &
0.0008 & 0.0036  & 
0.0016 & 0.0036  & 
0.0024 & 0.0029  \\ 

& CLAP-PT \cite{wu_largescale_2023}&
0.0020 & 0.0099 &
0.0041 & 0.0099 &
0.0065 & 0.0084 &
0.0013 & 0.0061  & 
0.0024 & 0.0054  & 
0.0038 & 0.0047  \\ 

& CLAP-FT \cite{wu_largescale_2023}&
\underline{0.0217}&	\underline{0.1047} &
\underline{0.0386} & \underline{0.0967}&
\underline{0.0682} & \underline{0.0877} &
0.0033 & 0.0125  & 
\underline{0.0067} & \underline{0.0131}  & 
\underline{0.0132} & \underline{0.0137}  \\ 

\cmidrule(lr){2-14}


& \textbf{LARP} (Ours)&
\textbf{0.0406} & \textbf{0.1964}&
\textbf{0.0753} & \textbf{0.1868}  & 
\textbf{0.1306} & \textbf{0.1669}  &
\textbf{0.0137} & \textbf{0.0536} &
\textbf{0.0233} & \textbf{0.0479} &
\textbf{0.0389} & \textbf{0.0446} \\

\midrule

\multirow{7.4}{*}{\textbf{DropoutNet \cite{volkovs_dropoutnet_2017}}} 

& Jukebox \cite{dhariwal_jukebox_2020}&
0.0022 & 0.0106 &
0.0034 & 0.0090 &
0.0063 & 0.0082 &
0.0008 & 0.0040 &
0.0014 & 0.0035 &
0.0023& 0.0030 \\

& MULE \cite{mccallum_supervised_2022}&
\underline{0.0286} & \underline{0.1464} &
\underline{0.0534} & \underline{0.1377} & 
\underline{0.0861} & \underline{0.1166} &
0.0066 & \underline{0.0261} &
\underline{0.0131} & \underline{0.0259} &
\underline{0.0236} & \underline{0.0257} \\

& CO-Playlist \cite{jimenez_pretraining_2023}& 
0.0016 & 0.0087 & 
0.0034 & 0.0088 &
0.0065 & 0.0083 &
0.0011 & 0.0053 & 
0.0017 & 0.0043 &
0.0030 & 0.0038 \\

& CLAP-PT \cite{wu_largescale_2023}&
0.0022 & 0.0114 &
0.0047 & 0.0117 &
0.0081 & 0.0105 &
0.0006 & 0.0031 &
0.0016 & 0.0034 &
0.0031 & 0.0034 \\

& CLAP-FT \cite{wu_largescale_2023} &
0.0283 & 0.1396 &
0.0531 & 0.1325 & 
0.0888 & 0.1158 &
\underline{0.0070} & 0.0259 &
0.0128 & 0.0247 &
0.0210 & 0.0232 \\

\cmidrule(lr){2-14}


& \textbf{LARP} (Ours)&
\textbf{0.0418} & \textbf{0.2171} &
\textbf{0.0784} & \textbf{0.2028} & 
\textbf{0.1375} & \textbf{0.1815} &
\textbf{0.0146} & \textbf{0.0558} &
\textbf{0.0245} & \textbf{0.0495} &
\textbf{0.0417} & \textbf{0.0471} \\

\midrule

\multirow{7.4}{*}{\textbf{CLCRec} \cite{wei_clcrec_2021}} 

& Jukebox \cite{dhariwal_jukebox_2020}&
0.0014 & 0.0078 &
0.0038 & 0.0092 & 
0.0071 & 0.0088 &
0.0010 & 0.0036 &
0.0013 & 0.0029 &
0.0021 & 0.0025 \\

& MULE \cite{mccallum_supervised_2022}&
\underline{0.0294} & \underline{0.1546} &
\underline{0.0521} & \underline{0.1391} &
\underline{0.0878} & \underline{0.1201} &
\underline{0.0087} & \underline{0.0324} &
\underline{0.0168} & \underline{0.0321} &
\underline{0.0310} & \underline{0.0326} \\

& CO-Playlist \cite{jimenez_pretraining_2023}& 
0.0026 & 0.0121 & 
0.0041 & 0.0103 &
0.0069 & 0.0091 &
0.0009 & 0.0060 & 
0.0015 & 0.0048 &
0.0023 & 0.0036 \\ 

& CLAP-PT \cite{wu_largescale_2023}&
0.0024 & 0.0111 &
0.0042 & 0.0103 & 
0.0071 & 0.0091 &
0.0012 & 0.0053 &
0.0017 & 0.0042 &
0.0030 & 0.0038 \\

& CLAP-FT \cite{wu_largescale_2023} &
0.0240 & 0.1137 &
0.0446 & 0.1088 & 
0.0814 &  0.1020 &
0.0076 & 0.0289 &
0.0141 & 0.0275 &
0.0241 & 0.0264 \\

\cmidrule(lr){2-14}


& \textbf{LARP} (Ours)&
\textbf{0.0416} & \textbf{0.2069} &
\textbf{0.0779} & \textbf{0.1962} & 
\textbf{0.1330} & \textbf{0.1737} &
\textbf{0.0151} & \textbf{0.0581} &
\textbf{0.0259} & \textbf{0.0527} &
\textbf{0.0423} & \textbf{0.0487} \\
\bottomrule
\end{tabular}
\label{table:baseline_comp}
}

\end{table*}

First, we compare our method with the various encoder benchmarks across all the downstream recommendation frameworks. As we can clearly see in Table \ref{table:baseline_comp}, our method consistently outperforms all the benchmarks across all metrics and their k values on both LFM and MPD. This performance shows the consistency of our approach over a variety of canonical metrics and cold-start music datasets. Furthermore, we can see that LARP significantly out performs the other relational representation learning benchmark, CO-Playlist \cite{jimenez_pretraining_2023}. This indicates the superiority of our relational learning paradigm and the value of adding textual inputs (which CO-Playlist does not use). Also, comparing with one of the two strongest benchmarks, MULE, we can see that even though MULE was trained on 1.8M examples of high-quality, proprietary music data \cite{mccallum_supervised_2022}, our method still performs better.

Second, we compare the performance of LARP, our method, across the various recommendation frameworks. Here, we can see that the difference between using a simple, non-parametrized recommendation model such as ItemKNN and a SOTA method such as DropoutNet or CLCRec is relatively insignificant. This indicates that LARP's exceptional performance can be achieved without reliance on a complex downstream recommendation model. This important finding highlights the benefits that the integration of cross-modal and relational signals can have in the cold-start setting. Furthermore, the lack of significant gains between various frameworks indicates that the key differentiating factor between these methods lies in the quality of the learned content representations. To us, this highlights the importance of considering the quality of content module as part of a cold-start recommendation task as compared to the downstream recommendation framework. 

\subsection{Ablation Study (RQ2)}

\setlength\dashlinedash{0.8pt}
\setlength\dashlinegap{1.5pt}
\setlength\arrayrulewidth{0.3pt}

\begin{table*} [h]
\centering
\caption{Ablation study with ItemKNN~\cite{sarwar_item_2001} as the recommender. The last row is our final method, LARP-TPC-\textit{fusion}. Each row bfore is an ablated version of LARP without (w/o) one of the contrastive modules.}


\label{table:ablation}
\resizebox{\linewidth}{!}{
\begin{tabular}{ll cccccc cccccc }
\toprule 

\multirow{2.3}{*}{\textbf{Baseline}} &
\multirow{2.3}{*}{\textbf{Encoder}} & 
\multicolumn{6}{c}{\textbf{MPD Dataset}} & 
\multicolumn{6}{c}{\textbf{LFM Dataset}} \\
\cmidrule(lr){3-8} \cmidrule(lr){9-14} && 
R@10 & N@10 & R@20 & N@20 & R@40 & N@40 & R@10 & N@10 & R@20 & N@20 & R@40 & N@40  \\
\midrule

Text &Bert \cite{devlin_pretraining_2019} &

0.0075 & 0.0415 &
0.0122 & 0.0351  & 
0.0191 & 0.0285 & 
0.0032 & 0.0128 & 
0.0052 & 0.0109  & 
0.0089 & 0.0104 \\ 

Audio  & HTS-AT \cite{chen_htsat_2022} &
0.0012 & 0.0072 &
0.0031 & 0.0079  & 
0.0071 & 0.0086  & 
0.0006 & 0.0025  & 
0.0011 & 0.0024 & 
0.0023 & 0.0025  \\ 


w/o $\ell^{\text{TTC}}$ &LARP-WTC& 
0.0256 & 0.1278 & 
0.0444 & 0.1168  & 
0.0779 & 0.1039  & 
0.0057 & 0.0232 & 
0.0104 & 0.0218  & 
0.0201 & 0.0218  \\

w/o $\ell^{\text{TPC}}$ & LARP-TTC& 
0.0373 & 0.1968 &
0.0724 & 0.1833  & 
0.1243 & 0.1631  &
\textbf{0.0141} & \textbf{0.0587} & 
\textbf{0.0256} & \textbf{0.0539}  & 
\textbf{0.0425} & \textbf{0.0496} \\

w/o \textit{fusion} & LARP-TPC& 
0.0357 & 0.1865 &
0.0641 & 0.1699  & 
0.1132 & 0.1517  & 
0.0094 & 0.0354  & 
0.0171 & 0.0333  & 
0.0286 & 0.0320  \\ 


LARP-\textit{fusion}& LARP-TPC-\textit{fusion}& 
\textbf{0.0406} & \textbf{0.1964}&
\textbf{0.0753} & \textbf{0.1868}  & 
\textbf{0.1306} & \textbf{0.1669}  &
0.0137 & 0.0536 &
0.0233 & 0.0479 &
0.0389 & 0.0446 \\


\bottomrule

\end{tabular}
 }

\end{table*}

To assess the effectiveness of each element in our proposed LARP architecture, we iteratively remove each component and evaluate the model's performance. In Table \ref{table:ablation} we present the results from each of the unimodal backbone encoders, followed by iterations of LARP in which we remove first the losses and then the \textit{fusion} layer. 

\begin{figure}[h]
\vspace{-.1in}
\centering
\begin{subfigure}[b]{0.45\textwidth}
    \includegraphics[width=\linewidth]{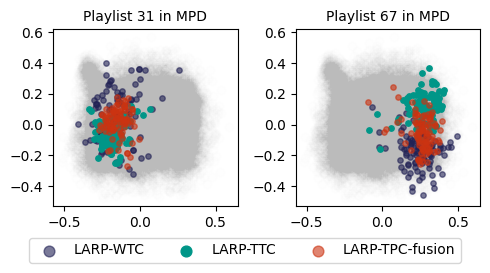}
   
\end{subfigure}
\begin{subfigure}[b]{0.45\textwidth}
    \includegraphics[width=\linewidth]{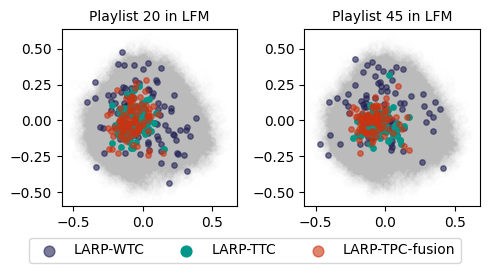}
\end{subfigure}
\caption{Within playlist alignment of tracks. Visualization of tracks associated with individual playlists. Note, gray cloud indicates the entire embedding set. }

\vspace{-.1in}
\label{fig:relational}
\end{figure}

First, we notice that each of the individual encoders has significantly lower performance than the overall architecture, motivating the need for the cross-attention layers and the WTC contrastive loss that is used to unify the uni-modal encoders. 

Second, we consider the fundamental architecture of LARP. Contrasting the performance of LARP-WTC with respect to its benchmark counterpart, ItemKNN/CLAP-FT in Table \ref{table:baseline_comp}, we can see that even without the remaining two losses, our method outperforms this comparable multi-modal benchmark by a significant margin. 

Third, we analyze the performance gains achieved by the integration of each loss. We can see that LARP-TTC, which begins to integrate relational content via track-track alignment, already shows noticeable improvement over the LARP-WTC which has no relational awareness. This is because the $\ell^{\text{TTC}}$ loss enables the model to account for acoustic or linguistic heterogeneity within the tracks of a playlist. For example, if we consider two tracks from different genres (e.g. Jazz and Country) that are both titled "Love", a method which is unaware of their relation in a playlist would be unaware of the unifying element (title) between them.

Fourth, we explore the importance of our fusion layer and its complementary $\ell^{\text{TPC}}$ loss. Simply the integration of a $\ell^{\text{TPC}}$ does not improve the performance and actually achieves worse performance than the $\ell^{\text{TTC}}$ loss. This is because the effectiveness of the $\ell^{\text{TPC}}$ loss hinges on the quality of the playlist representation that is used to anchor the children tracks. Without the self-attention layer, the model is unable to properly integrate information from a diverse set of seed tracks into a meaningful playlist representation. However, in adding this trained self-attention layer we are able to generate expressive playlist encodings that contribute to the final success of our LARP-TPC-\textit{fusion} method. 

Notably, we can see that there is a difference in which method performs best between the two datasets. For MPD, we can see that LARP-\textit{fusion} clearly outperforms LARP-TTC. We attribute this to the clear alignment between the MPD dataset setting (which was clearly designed for the playlist continuation task) and the formulation of the TPC loss that uses the \textit{fusion} encoder layer. Meanwhile, on LFM, we can see that the performance of LARP-TTC supersedes that of LARP-\textit{fusion}. This is due to the potential misalignment between the LFM dataset and the goals of the TPC loss. As we show in Table \ref{table:dataset_stats}, LFM shows noticeably higher levels of diversity in its tracks, thus it is possible that there is too much heterogeneity among the representational spaces of the tracks, making it difficult for a single attention layer to correctly capture a meaningful playlist representation. However, the flexibility of our approach allows for the selection of different variations that can be tailored to the needs of a dataset. 

Finally, we provide a case study in which we visualize the embeddings of various tracks in a playlist (see Figure \ref{fig:relational}) that are generated by each method presented in the ablation table results above (Table \ref{table:ablation}). To create these plots, we randomly select an example playlist, $p$, and its associated tracks, $S_p = \{s_i : s_i \in p\}$. We generate embeddings for the playlist, $e_p$, and tracks, $e_{S_p} = \{e_{s_i}: s_i \in S_p\}$ by Eqs. \eqref{eq:modal_avg} and  \eqref{eq:plist_gen}. We then apply 2-dimensional PCA on the set of embeddings and plot the outcomes. This visualization is significant in the context of the industrial setting where online recommendations are served by selecting nearest neighbours from a large scale embedding space representing an extremely large catalog. In this setting, efficiency can be gained from having similar songs tightly clustered in the overall embedding space. 
We show that the LARP-TPC-\textit{fusion} generates song embeddings that are most closely clustered together in the embedding space which is aligned with the purpose of the $\ell^{\text{TPC}}$ loss. By creating a playlist-track relational awareness in our representation learning module, we are able to identify clusters in unseen, held out evaluation tracks. Thus, LARP's fine-grained representation learning framework facilitates high performance from even a simple k-means recommendation framework like ItemKNN. 

\subsection{Model Study (RQ3)}
\subsubsection{Generalization Performance}

In addition to our evaluation of unseen playlists and items from within the same dataset, we also evaluate the generalization capabilities of LARP on completely different datasets. In Table \ref{table:generalization}, we showcase the performance of the best performing methods on the LFM dataset and evaluate against a LARP model which has been trained on the MPD dataset. Crucially, we can see that this generalization performance is incredibly strong, even going so far as to out perform the best benchmark performance on the LFM dataset. This is extremely important for the cold-start setting, because it shows that our model is robust to the dataset distribution - a gold standard in this setting.

\begin{table} [t]
\centering
\caption{Generalization results.  We apply LARP, pre-trained on MPD, and test its performance on LFM as compared with best performing benchmarks (both trained and tested on LFM). Best results in bold.}

\label{table:generalization}
\resizebox{\linewidth}{!}{
\begin{tabular}{l ccccccc }
\toprule 

\multirow{2.4}{*}{\textbf{Encoder}} & \multirow{2.4}{*}{\shortstack{\textbf{Training} \\ \textbf{Dataset}}}&\multicolumn{6}{c}{\textbf{LFM Results}} \\ 
 \cmidrule(lr){3-8} 
&&R@10 & N@10 & R@20 & N@20 & R@40 & N@40  \\
\midrule
 
MULE & LFM & 
0.0036 & 0.0140  & 
0.0066 & 0.0131  & 
0.0121 & 0.0133 \\ 
CLAP-FT & LFM & 
0.0033 & 0.0125  & 
0.0067 & 0.0131  & 
0.0132 & 0.0137  \\ 
LARP-TPC-\textit{fusion} &MPD& 
\textbf{0.0102} & \textbf{0.1311} &
\textbf{0.0165} & \textbf{0.1178}  & 
\textbf{0.0276} & \textbf{0.1079}  \\ 
\bottomrule
\end{tabular} }
\vspace{-.1in}
\end{table}

\subsubsection{Parameter Sensitivity}\label{sec:sensitivity}

As we show in the ablation study, one of the key components of our method's impressive performance is the integration of the TPC, or Track-Playlist Contrast, loss. Critical to the success of this loss function is a comprehensive methodology for generating meaningful playlist representations that can be used to properly anchor the various tracks within a playlist's collection. 

\begin{figure}[h]
    \centering
 \includegraphics[width=\linewidth]{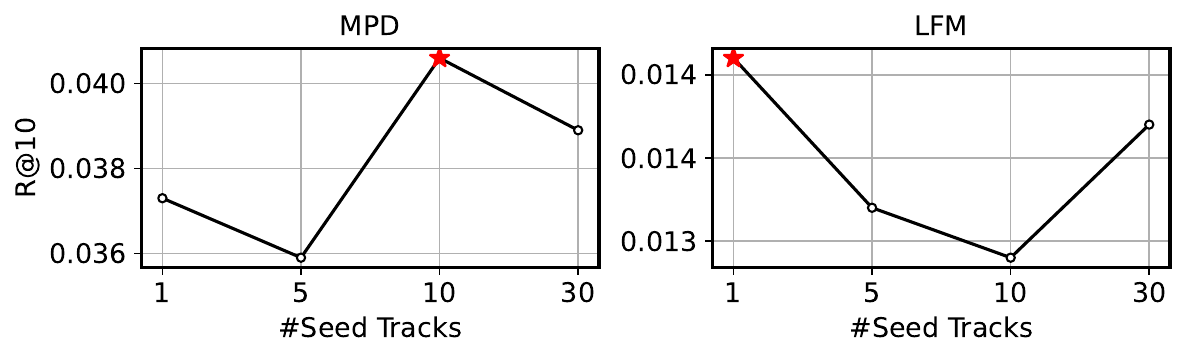}
    \caption{Sensitivity analysis. We show relationship between performance and number of seed tracks, $J$, for generating playlist representations. Note: the red star signifies the best results reported in the ablation tables above.}
    \label{fig:sensitivity}
\end{figure}

In our experimentation, we found that our model had some sensitivity to $J$, or the number of seed tracks which are used to compute the playlist representations during the training stage. And, most significantly, this parameter differs between datasets. As we show in Figure \ref{fig:sensitivity}, the best performance on the MPD dataset is achieved by using 10 seed tracks for the generation of the playlist representation by the \textit{fusion} playlist encoder (defined in Equation \ref{eq:fusion}). Meanwhile, in the LFM dataset, the best performance is achieved by using 1 or 30 seed tracks. We believe that this can be attributed to the difference in overall track diversity between the two datasets. As shown in Table \ref{table:dataset_stats}, LFM is generally less homogeneous than MPD in its overall track collection. This is expected since the LFM dataset is constructed over aggregate listening patterns while the MPD dataset is localized to specific playlists. Thus, these results can be attributed to the complexity of unifying a diverse set of tracks into a single playlist representation. In LFM, we can either control this complexity by significantly decreasing the number of seed tracks or providing a large enough subset that our method can find commonalities to anchor representations.

\subsubsection{Convergence Analysis}
\begin{figure}[t]
    \centering
  \includegraphics[width=\linewidth]{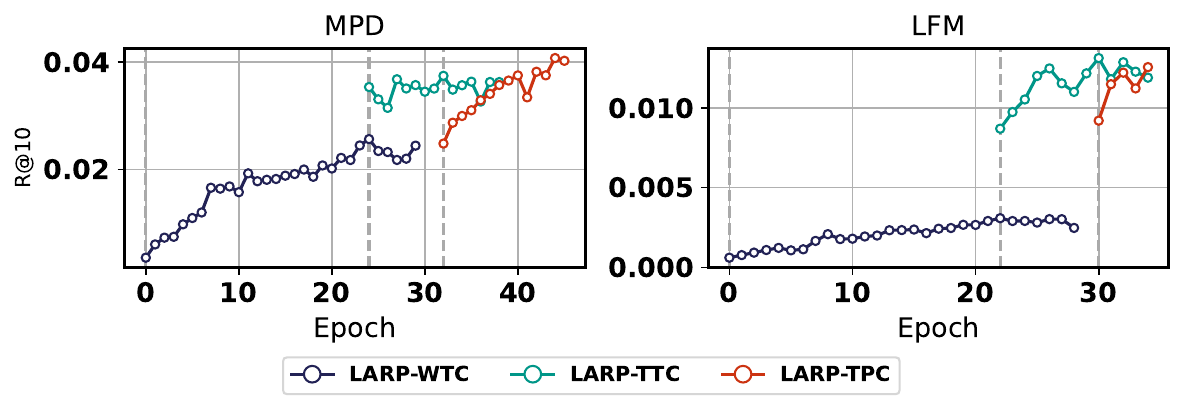}
    \caption{Convergence analysis of \method. Grey, vertical dashed line signifies the end of one training stage.}
    \label{fig:convergence}
    \vspace{-0.2in}
\end{figure}

The LARP model is trained in three consecutive stages, where each stage adds an additional loss to the training framework. In Figure \ref{fig:convergence}, we showcase the trajectory of performance that is yielded with the addition of each individual loss. The purpose of this plot is to showcase the significant improvements that are achieved with the integration of relational pre-training. Notably, we showcase the performance from the 5 epochs which follow the best performance to highlight that the optimal performance has, indeed been achieved. As we can see from these plots, on both datasets, there is a significant advantage to adding relational awareness. Interestingly, we can see that in the initial epochs of the $\ell_{TPC}$ training, there is a drop in performance. This can be attributed to the training of the self-attention layers, which are initialized with random values, and slowly tuned over the subsequent epochs of LARP-\textit{fusion} training. We note, that as mentioned in earlier sections, we can see that for MPD, which is specifically designed for the playlist continuation task, there are more advantages to the integration of TPC loss than for LFM. However, we believe this is expected given the high diversity of the LFM dataset. 

\section{Related Work}
In this section, we review related works to the \textit{cold start problem} and \textit{music understanding}.

\subsection{Cold-Start Recommendation} \label{sec:cold-start}
The issue of cold-start has been an important focus within the recommendation domain \cite{wenjie_EQUAL_23, barkan_cb2cf_2019, wei_clcrec_2021, li_from_2019, saveski_item_2014, vartak_meta_2017, volkovs_dropoutnet_2017, zhu_recommendation_2020, zhang_addressing_2014}. Many approaches in the field borrow from the domain of context based learning for handling partial or sparse inputs \cite{barkan_cb2cf_2019, zhang_addressing_2014}. Alternatively, many works randomly mask interactive signals \cite{volkovs_dropoutnet_2017, shi_adaptive_2019, zhu_recommendation_2020}. Finally, another branch of work uses content-based signals as a form of regularization or constraints during training \cite{saveski_item_2014, du_how_2020}. More recently, recent works have begun to integrate innovations from other tasks into the cold-start challenge such as contrastive learning \cite{wei_clcrec_2021}, meta-learning \cite{wu_m2eu_2023, dong_mamo_2020, lee_melu_2019}, and zero-shot or multimodal learning \cite{li_from_2019, ding_zero_2021, feng_zero_2021, szu_addressing_2016, ma_cirp_2024}. Crucially, these methods differ from our approach in that they are either using pre-trained representation learning methods, single modality content representations, or limit their cold-start formulation to include only cold-start users (while maintaining a shared item pool). 

Within the specific field of music recommendation, there have been many works that consider the topic of cold/warm start recommendation and playlist continuation tasks \cite{zamani_analysis_2019, volkovs_two_2018, yang_mmcf_2018, rubtsov_hybrid_2018, antenucci_artist_2018, briand_semi_2021, oord_deep_2013}. Notably, these works differ from ours in that they focus on either cold-start tracks \cite{oord_deep_2013, chou_addressing_2016}, users \cite{briand_semi_2021}, or playlists \cite{zamani_analysis_2019, volkovs_two_2018, yang_mmcf_2018, rubtsov_hybrid_2018, antenucci_artist_2018} but not both playlists and tracks (as our setting suggests). Furthermore, to our knowledge, there have yet to be any approaches that design methods that use both cross-modal and relational information to achieve their results.

\subsection{Music Understanding} \label{sec:mus_understanding}
In recent years, the music information retrieval community has used a variety of approaches to design music representation learning models. It has become a very standard practice within to apply state of the art vision models to images of the spectral content of audio samples using either convolutional neural networks \cite{kong_panns_2020, manco_learning_2022, park_representation_2018, pons_end--end_2018} or transformer-based architectures \cite{wu_largescale_2023, won_semisupervised_2021} to learn representations. In the last two years, innovations in vision-language tasks, have also found success when applied to audio-language tasks.  Most notably, \cite{huang_mulan_2022, wu_largescale_2023, manco_learning_2022, gardner_llark_2023, ma_leveraging_2024} have all borrowed architecture from contrastive learning methods used in vision-language tasks and applied them to audio by replacing the visual encoder with an audio backbone model. Several works have also proposed the integration of collaborative signals into their training paradigms \cite{huang_largescale_2020, park_representation_2018, jimenez_pretraining_2023} by using various collaboration-informed heuristics for selecting positive and negative pairs in their contrastive loss frameworks. Notably, our work differs from those presented in that none of these works integrate cross-modal \textit{and} relational signals without the use of collaborative filtering.

\section{Conclusion \& Future Work}

In this work, we proposed \method, a multi-modal relation-aware representation learning model for cold-start playlist continuation. At the core of our approach, we argued that the true bottleneck of cold-start learning lies in the quality of the content module. To this end, \method~ integrates multi-stage contrastive losses to learn multimodal and relation-aware item embeddings. First, we compared between the performance of \method~and a series of representation learning baseline methods, showing its superiority as an encoder. Second, we evaluated \method~in tandem with several canonical cold-start frameworks, showing that the quality of the embeddings produced by \method~significantly narrowed the gap between complex, parametrized recommendation frameworks (DropOutNet, CLCRec) and simple, unparameterized ones (ItemKNN). Finally, we demonstrated our method's generalization capabilities by performing cross-dataset evaluation.


We see many opportunities for future work. From the music representation perspective, we envision many expansions of the playlist encoding capacity of our model. Currently, the playlist encoder is not used during inference. With some refinement, it could become an independent playlist encoder during inference. Additionally, our current track-playlist loss function does not integrate sequential awareness, which, if included, could potentially improve performance. Finally, while the application of this work is centered around the music domain, we believe that the contributions and outcomes of this work have significance for the broader domains of multimodal representation learning and cold start recommendation. Here, we extract relational signal from the playlists, because they are formed via explicit user feedback and thus represent reliable relational signals. However, the intuition behind these loss functions is applicable to many other domains such as books, podcasts, or video and is thus a major contribution. The flexibility of our relational training paradigm has potential for expansion into a multi-headed encoder that generates album-level, playlist-level, or even user-level representations each with its own dimension of relational awareness. From the recommendation perspective, we envision further exploration into the warm-start problem, where we imagine that sparse interaction can be used to guide our relational training paradigm to further refine its representational qualities. Ultimately, we hope that our work demonstrates both the importance of content representations in recommendation and the power of encoding relational awareness into content modules.

\section*{Acknowledgements}
This research/project is supported by the National Research Foundation, Singapore under its Industry Alignment Fund – Pre-positioning (IAF-PP) Funding Initiative. Any opinions, findings and conclusions or recommendations expressed in this material are those of the author(s) and do not reflect the views of National Research Foundation, Singapore.

\bibliographystyle{ACM-Reference-Format}

\bibliography{main}
\balance

\appendix

\section{Detailed Loss Formulations}\label{ap:loss}

\noindent \textbf{(L1) WTC: Within-Track Contrastive Loss.}
As mentioned in \ref{sec:training_objective}, the L1 loss is formulated for cross-modal alignment within a track. As defined in Equation \ref{wtc}, this loss performs contrastive learning on the single-track level. Given a track, $s_i$, consisting of a language-audio pair, $(t_{s_i}, a_{s_i})$, we formally define this loss as: 
\begin{equation}
   \ell_{s_i}^{\text{WTC}} = \text{Contrast}(\mathbf{a}_{s_i}, \mathbf{t}_{s_i}).
\end{equation}
More specifically, the contrastive loss is formulated as:
\begin{equation}
     \text{Contrast}(\mathbf{a}_{s_i}, \mathbf{t}_{s_i}) =  \frac{1}{2} [\text{CE}(y_i^{\text{a2t}}, \hat{y}_i^{\text{a2t}}) + \text{CE}(y_i^{\text{t2a}}, \hat{y}_i^{\text{t2a}} )]
\end{equation} where softmax-normalized audio-to-text (a2t) and text-to-audio (t2a) similarities are defined by 
\begin{equation}
\begin{split}
     \hat{y}_i^{\text{a2t}} = \frac{\text{exp}(\operatorname{sim}(\mathbf{a}_{s_i}, \mathbf{t}_{s_i}) \mathbin{/} \tau)} {\sum_{j=1}^{N} \text{exp}(\operatorname{sim}(\textbf{a}_{s_i}, \textbf{t}_{s_n}) \mathbin{/} \tau)},\\
     \hat{y}_i^{\text{t2a}} = \frac{\text{exp}(\operatorname{sim}(\mathbf{t}_{s_i}, \mathbf{a}_{s_i}) \mathbin{/} \tau)}  {\sum_{j=1}^{N} \text{exp}(\operatorname{sim}(\textbf{t}_{s_i}, \textbf{a}_{s_n}) \mathbin{/} \tau)},
\end{split}
\end{equation}

such that sim($\cdot$) is the cosine similarity function, $\tau$ is a temperature hyper-parameter, $y_i^{\text{a2t}}, y_i^{\text{t2a}} \in \{0,1\}$ are one-hot ground-truth similarity vectors indicating the correct language-audio pairing, and $s_n$ is a negative sample, drawn from the momentum queue.

\noindent \textbf{(L2) TTC: Track-Track Contrastive Loss.} 
The second loss extends the formulation of the WTC (L1) loss to perform track-track contrastive learning. For two tracks $s_i = (t_{s_i}, a_{s_i})$ and $s_j = (t_{s_j}, a_{s_j})$, which share a parent playlist $p_k$, this loss is formulated as 

\begin{equation}
     \ell_{s_i,s_j}^{\text{TTC}} = \frac{1}{2} [\text{Contrast}(\textbf{a}_{s_i}, \textbf{t}_{s_j}) + \text{Contrast}(\textbf{t}_{s_i}, \textbf{a}_{s_j})].
\end{equation}\vspace{-.1in}Here, the each contrastive loss pairing is formulated as:
\begin{equation}
\begin{split}
\vspace{-.2in}
     \text{Contrast}(\mathbf{a}_{s_i}, \mathbf{t}_{s_j}) =  \frac{1}{2} [\text{CE}(y_{ij}^{\text{a2t}}, \hat{y}_{ij}^{\text{a2t}}) + \text{CE}(y_{ji}^{\text{t2a}}, \hat{y}_{ji}^{\text{t2a}} )], 
     \\
     \text{Contrast}(\mathbf{t}_{s_i}, \mathbf{a}_{s_j}) =  \frac{1}{2} [\text{CE}(y_{ij}^{\text{t2a}}, \hat{y}_{ij}^{\text{t2a}} ) + \text{CE}(y_{ji}^{\text{a2t}}, \hat{y}_{ji}^{\text{a2t}})],
\end{split}
\end{equation} where softmax-normalized audio-to-text (a2t) and text-to-audio (t2a) similarities are defined by 
\begin{equation}
\begin{split}
     \hat{y}_{ij}^{\text{a2t}} = \frac{\text{exp}(\operatorname{sim}(\mathbf{a}_{s_i}, \mathbf{t}_{s_j}) \mathbin{/} \tau)} {\sum_{n=1}^{N} \text{exp}(\operatorname{sim}(\textbf{a}_{s_i}, \textbf{t}_{s_n}) \mathbin{/} \tau)}, \\
     \hat{y}_{ji}^{\text{a2t}} = \frac{\text{exp}(\operatorname{sim}(\mathbf{a}_{s_j}, \mathbf{t}_{s_i}) \mathbin{/} \tau)} {\sum_{n=1}^{N} \text{exp}(\operatorname{sim}(\textbf{a}_{s_j}, \textbf{t}_{s_n}) \mathbin{/} \tau)}, \\
     \hat{y}_{ji}^{\text{t2a}} = \frac{\text{exp}(\operatorname{sim}(\mathbf{t}_{s_j}, \mathbf{a}_{s_i}) \mathbin{/} \tau)}  {\sum_{n=1}^{N} \text{exp}(\operatorname{sim}(\textbf{t}_{s_j}, \textbf{a}_{s_n}) \mathbin{/} \tau)}, \\ 
     \hat{y}_{ij}^{\text{t2a}} = \frac{\text{exp}(\operatorname{sim}(\mathbf{t}_{s_i}, \mathbf{a}_{s_j}) \mathbin{/} \tau)}  {\sum_{n=1}^{N} \text{exp}(\operatorname{sim}(\textbf{t}_{s_i}, \textbf{a}_{s_n}) \mathbin{/} \tau)}, 
\end{split}
\end{equation} such that sim($\cdot$) is the cosine similarity function, $\tau$ is a temperature hyper-parameter, and $s_n$ is a negative sample, drawn from the momentum queue. And the one-hot ground-truth similarity vectors indicating the correct language-audio pairing, are calculated as: 
\begin{equation}
\begin{split}
 y_{ij}^{\text{a2t}} = \Biggl\{ o \in \{ 0,1 \}^{B}: \sum_{v}^{B} o_v = O_{iv}  \Biggr\} , \\ 
 y_{ji}^{\text{a2t}} = \Biggl\{ o \in \{ 0,1 \}^{B}: \sum_{v}^{B} o_v = O_{jv}  \Biggr\} , \\ 
 y_{ij}^{\text{t2a}} = \Biggl\{ o \in \{ 0,1 \}^{B}: \sum_{v}^{B} o_v = O_{iv}  \Biggr\} , \\ 
 y_{ji}^{\text{t2a}} = \Biggl\{ o \in \{ 0,1 \}^{B}: \sum_{v}^{B} o_v = O_{jv}  \Biggr\} , \\ 
\end{split}
\end{equation} where $B$ is the batch size and $O_{N \times N}$ is the homogeneous, track-track co-occurrence graph defined in Section \ref{sec:prob_def}. 

\noindent \textbf{(L3) TPC: Track-Playlist Contrastive Loss.}
The third loss extends the formulation of TTC (L2) loss to perform track-playlist contrastive learning. For a track, $s_i$, its parent playlist, $p_k$, and a set of neighbour tracks that share the parent playlist, $J = \{s_j \in p_k \setminus s_i\}$, define this loss as: 
\begin{equation}
     \ell_{p_k,s_i}^{\text{TPC}} = \frac{1}{2} [\text{Contrast}(\textbf{a}_{p_k}, \textbf{t}_{s_i}) + \text{Contrast}(\textbf{t}_{p_k}, \textbf{a}_{s_i})],  
\end{equation} where the contrastive pairing is defined by 
\begin{equation}
\begin{split}
     \text{Contrast}(\mathbf{a}_{p_k}, \mathbf{t}_{s_i}) =  \frac{1}{2} [\text{CE}(y_{ki}^{\text{a2t}}, \hat{y}_{ki}^{\text{a2t}}) + \text{CE}(y_{ik}^{\text{t2a}}, \hat{y}_{ik}^{\text{t2a}} )], 
     \\
     \text{Contrast}(\mathbf{t}_{s_k}, \mathbf{a}_{s_i}) =  \frac{1}{2} [\text{CE}(y_{ki}^{\text{t2a}}, \hat{y}_{ki}^{\text{t2a}} ) + \text{CE}(y_{ik}^{\text{a2t}}, \hat{y}_{ik}^{\text{a2t}})],
\end{split}
\end{equation} 

where softmax-normalized audio-to-text (a2t) and text-to-audio (t2a) similarities are defined by 
\begin{equation}
\begin{split}
     \hat{y}_{ki}^{\text{a2t}} = \frac{\text{exp}(\operatorname{sim}(\mathbf{a}_{p_k}, \mathbf{t}_{s_i}) \mathbin{/} \tau)} {\sum_{n=1}^{N} \text{exp}(\operatorname{sim}(\textbf{a}_{p_k}, \textbf{t}_{s_n}) \mathbin{/} \tau)}, \\
     \hat{y}_{ik}^{\text{a2t}} = \frac{\text{exp}(\operatorname{sim}(\mathbf{a}_{s_i}, \mathbf{t}_{p_k}) \mathbin{/} \tau)} {\sum_{n=1}^{N} \text{exp}(\operatorname{sim}(\textbf{a}_{s_i}, \textbf{t}_{s_n}) \mathbin{/} \tau)}, \\
     \hat{y}_{ik}^{\text{t2a}} = \frac{\text{exp}(\operatorname{sim}(\mathbf{t}_{s_i}, \mathbf{a}_{p_k}) \mathbin{/} \tau)}  {\sum_{n=1}^{N} \text{exp}(\operatorname{sim}(\textbf{t}_{s_i}, \textbf{a}_{s_n}) \mathbin{/} \tau)}, \\ 
     \hat{y}_{ki}^{\text{t2a}} = \frac{\text{exp}(\operatorname{sim}(\mathbf{t}_{p_k}, \mathbf{a}_{s_i}) \mathbin{/} \tau)}  {\sum_{n=1}^{N} \text{exp}(\operatorname{sim}(\textbf{t}_{p_k}, \textbf{a}_{s_n}) \mathbin{/} \tau)}, 
\end{split}
\end{equation} where the playlist embeddings, $a_{p_k}, t_{p_k}$, are defined by the output of the \textit{fusion} layer defined in Section \ref{sec:training_objective} as: 
\begin{equation}
\begin{aligned}
    \textbf{a}_{p_k} &= \text{Self-Att}(\{\textbf{a}_{s_j}: s_j \in J \}), \\ \textbf{t}_{p_k} &= \text{Self-Att}(\{\textbf{t}_{s_j}: s_j \in J\}),
\end{aligned}
\end{equation} and that sim($\cdot$) is the cosine similarity function, $\tau$ is a temperature hyper-parameter, and $s_n$ is a negative sample, drawn from the momentum queue. 
More formally, if we consider $A_J = \{\textbf{a}_{s_j}: s_j \in J \}$ and $T_J = \{\textbf{t}_{s_j}: s_j \in J \}$ we can define this attention layer as: 
\begin{equation}
\begin{split}
    \text{Self-Att}(\textbf{A}_{J}) &=  \text{softmax}\Bigl(\frac{A_JW_Q^\top A_JW_K^\top}{\sqrt{d}}\Bigr) A_JW_V^\top \\ 
    \text{Self-Att}(\textbf{T}_{J}) &=  \text{softmax}\Bigl(\frac{T_JW_Q^\top T_JW_K^\top}{\sqrt{d}}\Bigr) T_JW_V^\top
\end{split}
\end{equation}
 where $W_Q, W_K, W_V\in \mathbb{R}^{|J|\times d}$, $d=256$. Meanwhile, the one-hot ground-truth similarity vectors indicating the correct language-audio pairing, are calculated as: 
\begin{equation}
\begin{split}
 y_{ki}^{\text{a2t}} = \Biggl\{ o \in \{ 0,1 \}^{B}: \sum_{v}^{B} r_v = R_{iv}  \Biggr\} , \\ 
 y_{ik}^{\text{a2t}} = \Biggl\{ o \in \{ 0,1 \}^{B}: \sum_{v}^{B} r_v = R_{jv}  \Biggr\} , \\ 
 y_{ik}^{\text{t2a}} = \Biggl\{ o \in \{ 0,1 \}^{B}: \sum_{v}^{B} r_v = R_{iv}  \Biggr\} , \\ 
 y_{ki}^{\text{t2a}} = \Biggl\{ o \in \{ 0,1 \}^{B}: \sum_{v}^{B} r_v = R_{jv}  \Biggr\} , \\ 
\end{split}
\end{equation} where $B$ is the batch size and $R_{M \times N}$ is the bipartite, playlist-track interaction graph defined in Section \ref{sec:prob_def}.

\vspace{-.1in}
\section{Cold-Start Implementations}\label{ap:hp_cold_start}

\subsection{ItemKNN \cite{sarwar_item_2001}} 
\subsubsection{Formulation.}
Given the pre-trained content features $\mathbf{E}_{\bar{\mathcal{S}}}$ for all tracks in testing set and the playlist representation $\mathbf{e}_{\bar{p}_q}$, we calculate the scores $\hat{y}_{\bar{p}_q}$ for the input partial playlist $\bar{p}_q$ with inner product to find held-out tracks. Formally:
\begin{equation}\label{eq:itemKNN_topk_similarities}
    \mathcal{\hat{S}} = \argmax_{\mathcal{\bar{S}}}(\hat{y}_{\bar{p}_q}),\quad 
    \hat{y}_{\bar{p}_q} = \mathbf{e}_{\bar{p}_q} \cdot \mathbf{E}_{\bar{S}}, 
\end{equation}
where $\mathcal{\hat{S}}$ are the predicted tracks. For the final recommendation, we keep the top-k highest tracks, thus selecting $|\mathcal{\hat{S}}|=K$. 

\subsection{DropoutNet. \cite{volkovs_dropoutnet_2017}}
\subsubsection{Formulation.} This method uses two stages of training.
\begin{itemize}[leftmargin=*]
    \item \textbf{WMF model}: Given a playlist-track interaction matrix $\mathbf{R}_{M\times N}$, WMF learns the latent collaborative embeddings, $\mathbf{Z}_{\mathcal{S}}$ and $\mathbf{Z}_{\mathcal{P}}$. 
    \item \textbf{DropoutNet model}: 
    Given the learned collaborative embeddings, $\mathbf{Z}_{\mathcal{S}}$ and $\mathbf{Z}_{\mathcal{P}}$, content features $\mathbf{E}_{\mathcal{S}}$ and $\mathbf{E}_{\mathcal{P}}$, generated from $\Phi(\cdot)$) DropoutNet concatenates them and passes trains DNN models, $f_p(\cdot)$ and $f_s(\cdot)$, with mean squared error (MSE) to recover these concealed entries and perform score prediction.
\end{itemize} 


During testing, where our cold-start setting precludes the possibility of calculating collaborative features, we follow the original paper's formulation to generate collaborative features by averaging them from the training set: 
\begin{equation}
    \tilde{\mathbf{z}}_{p} = \text{mean}(\{\mathbf{z}_{p}: p\in \mathcal{P}\}),\quad
    \tilde{\mathbf{z}}_{s} = \text{mean}(\{\mathbf{z}_{s}: s\in \mathcal{S}\}),\quad. 
\end{equation}
Thus, we input the partial playlist $\bar{p}_q$ in Equation~\ref{eq:DropoutNet_prediction} and  obtain the predicted scores $\hat{y}_{\bar{p}_q}$ for $\bar{p}_q$. Final predictions are generated using: 

\begin{equation}\label{eq:DropoutNet_prediction_testing}
    \hat{y}_{\bar{p}_q} = f_p(\mathbf{e}_{\bar{p}} || \tilde{\mathbf{z}}_{p})\cdot f_s(\mathbf{E}_{\mathcal{\bar{S}}} || \tilde{\mathbf{Z}}_{\mathcal{S}}), 
\end{equation}
where $\tilde{\mathbf{Z}}_{\mathcal{S}}$  is a matrix formed by row-wise stacking of $\tilde{\mathbf{z}}_{s}$ to match the shape requirement of concatenation with $\mathbf{E}_{\mathcal{\bar{S}}}$.
\subsubsection{Hyperparameter Settings.}
We set the learning rate to 0.005, L2 regularization to 0.1 and utilize the SGD optimizer with momentum of 0.9 for both WMF and DropoutNet training. The embedding size of WMF is 64 and the dimension of transformed features for DropoutNet is 256. Additionaly, we use three-layer MLPs to implement each of the transformation DNN models, $f_p$ and $f_s$, and a ratio of 0.2 for the feature dropout.
\vspace{-.1in}
\subsection{CLCRec \cite{wei_clcrec_2021}}
\subsubsection{Formulation.} Unlike DrpoutNet, CLCRec initializes the embeddings and learns them during the training procedure. 
Here, its training loss is:
\begin{equation}\label{}
    \ell_{\text{CLCRec}} = \text{Contrast}(\mathbf{z}_{p}, \mathbf{z}_{si}) + \text{Contrast}(f(\mathbf{e}_{si}), f(\mathbf{e}_{sj})), 
\end{equation}
where $f(\cdot)$ is a DNN model for transforming the pre-trained content features. CLCRec randomly chooses between the content features and collaborative embeddings using a predefined probability value as it learns to maximize the correlation between learned content features and latent playlist representations. The score prediction is similar with ItemKNN, but equipped with an additional DNN model, which can be denoted as:
\begin{equation}
    \hat{y}_{\bar{p}_q} = f(\mathbf{e}_{\bar{p}_q})\cdot f(\mathbf{E}_{\mathcal{\bar{S}}}).
\end{equation}


\subsubsection{Hyperparameter Settings.}
We tune the learning rate to 0.001, L2 regularization to 0.1 and use the Adam optimizer. The temperature for contrastive learning is 2.0 and we set the probability of replacement to 0.5.





\end{document}